\documentclass{ws-ijmpe}
\usepackage[super,compress]{cite}
\usepackage{epsfig}
\usepackage{epstopdf}
\begin{document}

\markboth{Jameel-Un Nabi and Muhammad Majid}{Gamow-Teller strength
and lepton captures rates on $\mathbf{^{66-71}}$Ni in stellar
matter}

\catchline{}{}{}{}{}

\title{\textbf{Gamow-Teller strength and lepton captures rates on $\mathbf{^{66-71}}$Ni in
stellar matter}}

\author{Jameel-Un Nabi}

\address{Faculty of Engineering Sciences,\\Ghulam Ishaq Khan (GIK) Institute of
Engineering Sciences and Technology, Topi 23640, Khyber Pakhtunkhwa,
Pakistan\\ jameel@giki.edu.pk}

\author{Muhammad Majid}

\address{Faculty of Engineering Sciences,\\GIK Institute of
Engineering Sciences and Technology, Topi 23640, Khyber Pakhtunkhwa,
Pakistan\\ majid.phys@gmail.com}

\maketitle

\begin{history}
\received{Day Month Year}
\revised{Day Month Year}
\end{history}

\begin{abstract}
Charge-changing transitions play a significant role in stellar
weak-decay processes. The fate of the massive stars is decided by
these weak-decay rates including lepton (positron and electron)
captures rates, which play a consequential role in the dynamics of
core collapse. As per previous simulation results, weak interaction
rates on nickel isotopes have significant influence on the stellar
core vis-$\grave{a}$-vis controlling the lepton content of stellar
matter throughout the silicon shell burning phases of high mass
stars up to the presupernova stages. In this paper we perform a
microscopic calculation of Gamow-Teller charge-changing transitions,
in the $\beta$-decay and electron capture directions, for
neutron-rich nickel isotopes ($^{66-71}$Ni). We further compute the
associated weak-decay rates for these selected nickel isotopes in
stellar environment. The computations are accomplished by employing
the deformed proton-neutron quasiparticle random phase approximation
(pn-QRPA) model. A recent study showed that the deformed pn-QRPA
theory is well suited for the estimation of Gamow-Teller
transitions. The astral weak-decay rates are determined over
densities in the range of 10 -- 10$^{11}$g/cm$^{3}$ and temperatures
in the range of 0.01$\times$10$^{9}$ -- 30$\times$10$^{9}$K. The
calculated lepton capture rates  are compared with the previous
calculation of Pruet and Fuller. The overall comparison demonstrates
that, at low stellar densities and high temperatures, our electron
captures  rates are bigger by as much as two orders of magnitude.
Our results show that, at higher temperatures, the lepton capture
rates are the dominant mode for the stellar weak rates and the
corresponding lepton emission rates may be neglected.
\end{abstract}


\ccode{PACS numbers: 23.40.Bw, 23.40.-s, 26.30.Jk, 26.50.+x,
97.10.Cv}


\section{Introduction}
Majority of stellar processes evolve via nuclear reactions involving
a large number of exotic nuclides. Weak-decay properties of these
nuclei play a decisive part in astrophysical phenomena. Significant
efforts have been developed in last decades to calculate the masses
and other nuclear properties of neutron-rich nuclei at radioactive
nuclear beam facilities around the globe. However, most of these
exotic nuclei cannot be synthesized in laboratories and theoretical
estimates of weak-decay properties became more demanding to help us
understand these stellar processes. The weak rates are the crucial
constituents to be identified in essentially all astrophysical
processes \cite{Bur59}. Stellar evolution and the associated
nucleosynthesis process have been the attention of much calculation
\cite{Bur59,Arn96}. At the later stages of burning cycle of massive
stars, an iron core is eventually created and then no further
nuclear fuel is available for ignition of a new cycle. The stellar
core gradually becomes unstable and collapses due to the capture of
free electrons and photodisintegration processes. The collapse is
sensitive to the electron to-baryon ratio (Y$_{e}$) and to the core
entropy \cite{Bet79}. These parameters are governed by
charge-changing transition processes, i.e. $\beta$-decay and
electron (positron) captures. The core collapse simulation greatly
depends on the electron capture (EC) of heavy nuclei \cite{Hix03}.
The $fp$-shell nuclei capture the electrons due to which the Y$_{e}$
is reduced in the initial phase of collapse. The massive star late
evolutionary phases are significantly affected by charge-changing
transition processes which in turn control the stellar core entropy
and Y$_{e}$ of the presupernova star. These weak interaction
reactions also control the Chandrasekhar mass (M$_{ch}$), related to
Y$_e$$^{2}$ \cite{Cha39}. EC reduces the quantity or number of
electrons that are present for pressure support, while the
$\beta$-decay increase the quantity of these electrons. In both
processes, (anti)neutrinos are produced which, for stellar density,
$\rho \leq$ 10$^{11}$ gcm$^{-3}$, outflow from the stars carrying
away entropy and energy from the core. At high density and
temperature, heavy nuclei capture the electrons inside the stars
thereby reducing the degeneracy pressure, and lead to the
neutronization of stars. The EC importance for the collapse of
presupernova  stars were discussed in Ref. \cite{Bet90}. The
positron captures (PC) are of crucial significance in astrophysical
core, particularly in low density and high temperature regions. In
these circumstances, a slightly higher positron (e$^{+}$)
concentration is produced via e$^{-}$+e$^{+}$ $\leftrightarrow$
$\gamma+\gamma$ equilibrium state that favors the electron(positron)
pairs. The race (and possibly the equilibrium) among PC and EC is a
central constituent for the Type-II supernovae modeling (see e.g.
Ref. \cite{Nab07a}).

During the final evolutionary processes of high mass stars, EC (PC)
and stellar $\beta$-weak rates are controlled by  Gamow-Teller (GT)
transitions (and to a much lesser extent by Fermi transitions). The
GT  transitions are of spin-isospin ($\sigma\tau$) type and play a
significant role in nuclear weak-decay reactions. These spin-isospin
transitions are not only crucial in the domain of nuclear physics,
but also in many astrophysical phenomena (including stellar
evolution and associated nucleosynthesis process). The PC (EC) weak
rates are determined by the distributions of GT$_{-}$ (GT$_{+}$)
transition strength. Protons are transformed into neutrons in the
GT$_{+}$ transition while the GT$_{-}$ strengths are responsible for
transforming neutrons into protons. The total GT$_{+}$
charge-changing strength is related to the EC cross sections
\cite{Ron93}. Specially for the \emph{fp}-shell nuclei, the GT
transition strengths are of fundamental importance for supernova
physics \cite{ffn80}. In stellar environment the nuclides are
completely ionized and so only continuum EC is possible  from the
degenerate electron plasma. GT$_{+}$ strength on nuclei with (A = 50
-- 65) were measured primarily through (n,p) reactions at forward
angles \cite{Ron93,Vet89,Kat94}. In contrast the GT$_{-}$ transition
strengths were measured using the (p,n) reactions
\cite{Vet89,Rap83,And90}. These experiments demonstrated that the
overall GT transitions were quenched (as compared to the results of
independent particle theory). Further it was also reveled that the
GT strength was fragmented and spread over several final energy
levels in the daughter. It was emphasized that the residual
interactions initiated these effects amongst the valence nucleons
and an exact explanation of such correlations were deemed necessary
for a precise determination of the astrophysical interaction
processes. The first ever extensive work for the estimation of
astrophysical weak-decay rates was done by Fuller, Fowler and
Newmann (FFN) using the independent particle model \cite{ffn80}.
Their  work comprised of calculation of weak rates (including
positron/electron capture, positron/electron emission and
(anti)neutrino energy loss rates) for free nucleons and 226 nuclides
with mass range 21 to 60. All the measured data present at that time
were incorporated by the authors. A zero-order shell model was
employed for the estimation of excitation energies and GT strength
distributions. Aufderheide and collaborators \cite{auf94} further
emphasized the  significance of weak interaction rates in the iron
stellar core and extended the FFN rates for heavier nuclei with mass
greater than sixty. Experimental data \cite{Ron93, Kat94, And90,
Rap83} later revealed the GT centroid misplacement used in the FFN
\cite{ffn80} and Aufderheide et al. \cite{auf94} calculations. This
triggered efforts for a microscopic calculation of astrophysical
weak interaction rates. The large-scale shell model (LSSM)
\cite{lan00} and the pn-QRPA model \cite{nab99,nab04} are the two
most effective and widely employed models that are used for the
accurate and microscopic computations of weak interaction processes.
The pn-QRPA model has two main advantages. It can be used for the
calculation of weak rates for any arbitrarily heavy nucleus. The
pn-QRPA model calculates ground and excited states GT strength
functions in a microscopic fashion and does not use the Brink-Axel
hypothesis \cite{Bri55} for determination of excited state GT
strength functions (as used in LSSM and FFN calculations). This
feature of the pn-QRPA model increases its utility for the
calculation of weak rates in stellar environment where there is a
finite probability of occupation of parent excited states. These
excited states contribute substantially to the total weak rates and
a microscopic calculation of weak rates from parent excited states
surely adds to the reliability of the total rates. Recent
calculations \cite{Nab16,Joh15,Mis14} have shown that for a reliable
and accurate estimation of stellar decay rates the Brink-Axel
hypothesis is a poor approximation.

The pn-QRPA model has been used successfully with various
interactions and variations by many authors in the past. The
continuum QRPA approach was earlier used by Borzov \cite{Bor06}
which was based on the self-consistent ground state description in
the framework of the density functional theory. Sarriguren
\cite{sar13} used the QRPA model with a Skyrme Hartree-Fock
self-consistent mean field for calculation of stellar EC rates. The
finite-amplitude method was used by Mustonen and Engel \cite{mus16}
to perform a QRPA calculation of $\beta$-decay in even-even
axially-deformed nuclei with Skyrme energy-density functional. The
QRPA equations were also solved using the so-called Pyatov formalism
based on the definition of the effective Hamiltonian which
considered Dirac restrictions. Kuliev and collaborators
\cite{Kul00}, Babacan et. al. \cite{Bab04,Bab05} used this method to
calculate scissor mode vibrations, isobar analog states (IAS) and
isospin admixtures in the ground states of spherical nuclei as well
as GT Resonance (GTR) states. Few authors also used a hybrid model
scheme where allowed GT transitions were calculated within the QRPA
formalism and gross theory for first-forbidden transitions
\cite{mol03}.

The deformed pn-QRPA theory is a reliable approach, microscopic in
nature, to generate the GT transition functions. These strength
functions establish a non-trivial and primary influence to the decay
and capture processes amongst the iron mass range nuclides. It was
Nabi and Klapdor-Kleingrothaus, who employed the theory of pn-QRPA
for the first time, for the calculation of weak-decay processes on a
widespread range of density and temperature  for \emph{sd}-
\cite{nab99} and \emph{fp/fpg}-shell nuclei \cite{nab04} in stellar
environment. These global calculations were later improved, on
case-to-case basis (e.g. \cite{nab07, nab08}), with the usage of
further proficient algorithms, refinement of model parameters,
integration of experimental values and newest data from mass
compilations. The reliability and accuracy of the deformed pn-QRPA
model calculation were discussed in Ref.\cite{nab04}. Recently six
different pn-QRPA models were considered, for the calculation of GT
transitions in chromium isotopes \cite{sad15}. It was concluded in
Ref. \cite{sad15} that the current deformed pn-QRPA model reproduced
well the existing experimental data and possessed decent predictive
power to estimate the half-lives for unknown nuclei, compared to
other five pn-QRPA models discussed in the study.

During the evolution of cores of high mass stars, weak-rates on
nickel isotopes are considered to play a consequential role in the
evolutionary process. Several simulation results of massive stars
showed that EC and $\beta$-decay rates due to nickel isotopes
significantly alter the Y$_{e}$ of the stellar core (for detail
refer to \cite{auf94,Heg01}). The GT strength distribution and EC on
$^{56}$Ni, employing the pn-QRPA model, was first reported in Ref.
\cite{nab05}. The calculations were later extended to heavier
isotopes of nickel $^{57-65}$Ni \cite{nab12}. The GT and unique
first-forbidden  $\beta$-decay rates for heavy nickel isotopes with
mass range 72 to 78, employing the same nuclear model, was later
calculated by \cite{nab14}. First-forbidden transitions (including
rank 0, rank 1 and rank 2 contributions) for even-even nickel
isotopes $^{72,74,76,78}$Ni using different QRPA methods and related
$\beta$-decay properties were recently reported \cite{nab16a}. There
was a need to determine the GT transitions and  associated stellar
weak-interaction rates for  neutron-rich nickel isotopes
$^{66-71}$Ni. In this work, we calculate GT distribution functions
and the related lepton capture rates on neutron-rich Ni nuclides
with mass range A = 66 -- 71, using the pn-QRPA model.

The manuscript is designed as follows. The theoretical formalism
employed in the computation of stellar weak processes is discussed
briefly in Sec.~2. Discussion on calculated weak rates  and
comparison with previous calculation are shown in Sec.~3. The
conclusion is finally presented in Sec.~4.

\section{Formalism}
For the pn-QRPA calculation the Hamiltonian selected is of the form
\begin{equation} \label{GrindEQ__1_}
H^{QRPA} =H^{SP} + V^{pairing} + V_{GT}^{pp} + V_{GT}^{ph},
\end{equation}
where $H^{SP}$ represents the Hamiltonian of single particle,
$V^{pairing} $ represents the pairing force (within the BCS
approximation the pairing between nucleons is treated ),
$V_{GT}^{pp}$ represents the particle-particle ($pp$) and
$V_{GT}^{ph}$ is the particle-hole ($ph$) Gamow-Teller forces.
Single particle (SP) wave functions and their energies were
estimated using Nilsson model \cite{nil55}.   The $ph$ and $pp$
interaction constants were characterized by $\chi$ and $\kappa$,
respectively. Selection of $ph$ and $pp$ parameters were done in an
optimum way to reproduce the available experimental half-lives and
to satisfy the Ikeda sum rule (ISR) \cite{isr63}. In this work, we
took $\chi$ = 4.2/A (MeV), displaying a $1/A$ dependence
\cite{Hom96} while $\kappa$ was fixed at 0.10 MeV for all isotopes.
Other parameters essential for capture rates calculations were the
nuclear deformations ($\beta_{2}$), the pairing gaps ($\Delta _{n}$
and $\Delta _{p}$), the parameters for Nilsson potential (NP) and
the Q-values. The NP parameters were chosen from Ref. \cite{rag84}
and $\hbar \omega = 41/A^{1/3}$ MeV was considered as the Nilsson
oscillator constant, similar for neutrons and protons. The estimated
half-lives (T$_{1/2}$) values weakly rely on the $\Delta _{p}$ and
$\Delta _{n}$ \cite{hir91}. Thus, the conventional values of
\begin{equation}
\Delta _{n} =\Delta _{p} =12/\sqrt{A}  (MeV),
\end{equation}
were taken in this paper. Nuclear deformation parameter $\beta_{2}$
was determined using
\begin{equation}
\beta_{2} = \frac {125 (Q_{2})} {1.44 (A)^{2/3} (Z)},
\end{equation}
where $Q_{2}$ denote the electric quadrupole moment chosen from
\cite{Moe81}. We also checked our calculation with $\beta_{2}$
values taken directly from Ref. \cite{Moe95} and our end results
changed by less than 1$\%$. It is to be noted that the nickel
isotopes considered in this work are essentially spherical with very
small $\beta_{2}$ values and in order to be consistent with our
previous calculations we decided to use deformation values
calculated from Ref. \cite{Moe81}. Q-values were chosen from the
latest mass compilation of \cite{aud12}.

The charge changing transitions, in pn-QRPA model are defined in
terms of phonon creation operators. The proton-neutron QRPA phonons
are expressed as
\begin{equation}
A_{\omega}^{+}(\mu)=\sum_{pn}(X^{pn}_{\omega}(\mu)a_{p}^{+}a_{\bar{n}}^{+}-Y_{\omega}^{pn}(\mu)a_{n}
a_{\bar{p}}).
\end{equation}
In Eq.~(4) the summation runs over all proton-neutron pairs having
$\mu$ = \textit{m$_{p}$-m$_{n}$} = -1, 0, 1, where
\textit{m$_{p}$}(\textit{m$_{n}$}) represent the third component of
angular momentum of proton (neutron). The \textit{a$^{+}_{p(n)}$}
represent the creation operator of a quasi-particle (q.p) state of
proton (neutron), whereas \textit{$\bar{p}$} denotes the time
reversed state of \textit{p}. The ground level of the theory with
respect to the QRPA phonons is expressed as the vacuum,
A$_{\omega}(\mu)|QRPA\rangle$ = 0. The phonon operator
A$_{\omega}^{+}(\mu)$, having excitation energy ($\omega$) and
amplitudes (\textit{X$_{\omega}, Y_{\omega}$}) are achieved by
solving the renowned RPA matrix equation.
\begin{equation}
\begin{bmatrix}
A & B  \\
-B & -A
\end{bmatrix}
\begin{bmatrix}
X \\
Y
\end{bmatrix}
= \omega
\begin{bmatrix}
X \\
Y
\end{bmatrix},
\end{equation}
where \textit {X} and \textit {Y} are forward- and backward-going
amplitudes respectively. The $\omega$ denotes the energy eigenvalues
and the two sub-matrices are specified as
\begin{equation}
\begin{split}
A_{pn,p'n'}=\delta(pn,p'n')(\varepsilon_{p}+\varepsilon_{n})+V_{pn,p'n'}^{pp}(u_{p}u_{n}u_{p'}u_{n'}+v_{p}v_{n}v_{p'}v_{n'})\\
+V_{pn',p'n'}^{ph}(u_{p}v_{n}u_{p'}u_{n'}+v_{p}u_{n}v_{p'}v_{n'}),
\end{split}
\end{equation}
\begin{equation}
\begin{split}
B_{pn,p'n'}=+V_{pn,p'n'}^{pp}(u_{p}u_{n}u_{p'}u_{n'}+v_{p}v_{n}v_{p'}v_{n'})\\
-V_{pn',p'n'}^{ph}(u_{p}v_{n}u_{p'}u_{n'}+v_{p}u_{n}v_{p'}v_{n'}),
\end{split}
\end{equation}
where the $\varepsilon_{p(n)}$ shows the q.p energies of
proton(neutron), whereas  \textit{v$_{p(n)}$/u$_{p(n)}$} denote the
occupation/unoccupation amplitudes and were achieved performing a
BCS calculation. Detailed solution of RPA matrix equation can be
seen in Refs. \cite{mut89, hir91}.

Capture rates on $^{66-71}$Ni isotopes were calculated for the following two processes:\\
i. Electron capture (EC)
\begin{equation}
 ^{(66-71)}Ni + e^{-} \longrightarrow ^{(66-71)}Co + \nu
\end{equation}
 ii. Positron capture (PC)
\begin{equation}
 ^{(66-71)}Ni + e^{+} \longrightarrow  ^{(66-71)}Cu + \bar {\nu}.
\end{equation}

The EC and PC rates from the \emph{nth} level of parent nuclide
transition to the \emph{mth} level of daughter nuclide is specified
as
\begin{equation}
 \lambda _{nm}^{EC(PC)} =\ln 2\frac{f_{nm}(T,\rho
 ,E_{f})}{(ft)_{nm}},
\end{equation}
the term $(ft)_{nm}$ is linked to the reduced transition probability
($B_{nm}$) by
\begin{equation}
(ft)_{nm} =D/B_{nm},
\end{equation}
the value of D is considered constant and is taken as
\begin{equation}
 D=\frac{2\ln 2\hbar ^{7}\pi^{3}}{g_{v}^{2} m_{e}^{5}
c^{4} }.
\end{equation}
The reduced transition probabilities $B_{nm}$'s are given by
\begin{equation}
 B_{nm}=((g_{A}/g_{V})^{2} B(GT)_{nm}) + B(F)_{nm}.
\end{equation}
The value of D is chosen as 6295 s as in Ref. \cite{hir93} and
$g_{A}/g_{V}$ as -1.254. The reduced Fermi and GT transition
probabilities are specified by
\begin{equation}
B(F)_{nm} = \frac{1}{2J_{n} +1} \langle{m}\parallel\sum\limits_{k}
\tau_{\pm}^{k}\parallel {n}\rangle|^{2}
\end{equation}
\begin{equation}
B(GT)_{nm} = \frac{1}{2J_{n} +1} \langle{m}\parallel\sum\limits_{k}
\tau_{\pm}^{k}\overrightarrow{\sigma}^{k}\parallel {n}\rangle|^{2},
\end{equation}
here $\overrightarrow{\sigma}(k)$ and and $\tau_{\pm }^{k}$
represent the spin and the isospin operators, respectively.

The excited states in the pn-QRPA model can be constructed as
phonon-correlated multi-quasi-particle states. The RPA is formulated
for excitations from the $J^{\pi} = 0^{+}$ ground state of an
even-even nucleus. The model extended to include the quasiparticle
transition degrees of freedom yields decay half-lives of odd-mass
and odd-odd parent nuclei with the same quality of agreement with
experiment as for even-even nuclei (where only QRPA phonons
contribute to the decays) \cite{mut92}. For  the odd-A nickel
isotopes, the ground state can be expressed as a one-quasiparticle
state, in which the odd quasiparticle (q.p) occupies the single-q.p.
orbit of the smallest energy.  Then there exits two different type
of transitions: phonon transitions with the odd neutron acting only
as a spectator and transition of the odd neutron itself. For the
later case, phonon correlations were introduced to one-quasiparticle
states in first-order perturbation \cite{mut89}. When a nucleus has
an odd neutron, some low-lying states are obtained by lifting the
q.p. in the orbit of the smallest energy to higher-lying orbits \cite{mut92}. For odd-A nickel isotopes, the excited states were constructed,\\
(1) by lifting the odd neutron from ground state to excited states (one-q.p. state),\\
(2) by three-neutron states, corresponding to excitation of a neutron (three-q.p. states), or,\\
(3) by one-neutron two-proton states, corresponding to excitation of
a proton (three-q.p. states).

Similarly the low-lying states of an odd-proton even-neutron nucleus (daughter nucleus) were constructed,\\
(1) by exciting the odd proton from ground state (one-q.p. states),\\
(2) by three-proton states, corresponding to excitation of a proton (three-q.p. states), or,\\
(3) by one-proton two-neutron states, corresponding to excitation of
a neutron (three-q.p. states).

The transition amplitudes between the multi-quasi-particle states
can be reduced to those of single-quasi-particle states and can be
seen from Ref. \cite{nab99}. They are not reproduced here because of
space consideration.

The phase space ($f_{nm}$) integral was taken over total energy.
Adopting natural units $(\hbar=c=m_{e}=1)$, it is given by
\begin{equation}
f_{nm} = \int _{w_{1} }^{\infty }w\sqrt{w^{2} -1}(w_{m} +w)^{2}
F(\pm Z, w) G_{\mp }dw,
\end{equation}
(for positron capture (PC) lower sign is used while in case of
electron capture (EC) upper sign is used),

and the phase space integrals for positron (lower sign) or electron
(upper sign) emission is given by
\begin{equation}
f_{nm}=\int _{1}^{w_{m} }w\sqrt{w^{2} -1}(w_{m} -w)^{2}F(\pm Z,w)(1-
G_{\mp })dw.
\end{equation}

In Eqs.~(16) and (17), $w$ is the total electron kinetic energy
(K.E) including its rest mass, whereas the total capture threshold
energy for EC and PC is denoted by $w_{l}$. G$_{+}$ and G$_{-}$
denote the positron and electron distribution functions,
respectively, given as
\begin{equation}
G_{+} =\left[\exp \left(\frac{E+2+E_{f} }{kT} \right)+1\right]^{-1}
\end{equation}

\begin{equation}
 G_{-} =\left[\exp \left(\frac{E-E_{f} }{kT} \right)+1\right]^{-1},
\end{equation}
where E = (w - 1) represents the of electron K.E, $E_{f}$ denote the
electrons Fermi energy, T is the temperature and $k$ is the
Boltzmann constant. The Fermi functions denoted by F (Z, w) are
determined by using the same method as in Ref. \cite{gov71}. It is
to noted that if the  positron (or electron) emission energy
($w_{m}$) value is larger than -1, then $w_{l}$ is equal to 1, and
if $w_{m}$ $\leq$ 1, then $w_{m}=|w_{l}|$, where $w_{m}$ denote the
energy of total $\beta$-decay, and it is given as
\begin{equation} w_{m} = E_{n} -E_{m} + m_{p} -m_{d},
\end{equation}
\noindent where $E_{n}$ ($E_{m}$) and $m_{p}$ ($m_{d}$) represent
the excitation energies and masses of the parent (daughter) nuclide,
respectively. As the temperature inside the stars is very high so
there exist a finite possibility of occupation of parent excited
levels. The total lepton captures rates are specified by
\begin{equation}
\lambda^{EC(PC)} =\sum _{nm}P_{n} \lambda _{nm}^{EC(PC)},
\end{equation}
where $P_{n}$ obeys normal Boltzmann distribution. In Eq. 21, the
summation is taken over all final and initial levels until
reasonable convergence is obtained in the calculated rates.

\section{Results and Discussion}
Nickel (Ni) isotopes are considered to play a fundamental role in
the presupernova evolution of high mass stars. Previous simulation
results show that Ni isotopes ($^{56-61}$Ni, $^{63-69}$Ni and
$^{71}$Ni) significantly alter the fraction of electron to-baryon
inside the stellar core of massive stars \cite{auf94}. The GT
strength and weak rates on $^{56-65}$Ni isotopes have been discussed
in detail previously (see  \cite{nab05,nab08a,nab12}). In this paper
we would like to discuss the calculated GT$_{\pm}$ transition

functions, terrestrial half-lives (T$_{1/2}$) and stellar lepton
capture rates, on heavy Ni nuclides with mass ranging from A = 66 to
71. Our computed pn-QRPA results were quenched.  We used a quenching
factor ($f_{q}$) of 0.6 (similar to other microscopic calculations
including shell model) \cite{Vet89, Gaa83}. Authors in Ref.
\cite{Vet89} and \cite{Ron93} suggested a similar $f_{q}$ value of
0.6 for the RPA calculations for $^{54}$Fe nuclei, when matching
their measured strengths to RPA results. The total GT$_{\pm}$
strengths are linked to the re-normalized Ikeda sum rule
(ISR$_{re-norm}$) as
\begin{equation}
ISR_{re-norm} = \sum B(GT_{-}) - \sum B(GT_{+})\cong
3f_{q}^{2}(N-Z). \label{Eqt. ISR}
\end{equation}

Table~1 show that our calculated ISR$_{re-norm}$ values are in
excellent agreement with the theoretical prediction.

The GT charge-changing strength distribution for the ground level of
selected $^{66-71}$Ni isotopes are shown in Figs.~1 and 2. The
charge-changing transitions for excited states are not presented
because of space consideration. The ASCII files of both (ground and
excited level) GT distribution functions may be requested from the
corresponding author. We considered around 100 states in daughter
nuclei, up to excitation energy $\sim$ 45 MeV both for positron
capture (PC) and electron capture (EC), respectively. GT transitions
are central mode of excitation for the capture rates and
$\beta$-decay throughout the presupernova evolution. Figs.~1 and 2
show the B(GT$_{\pm}$) strength functions for the selected
$^{66-71}$Ni in GT$_{-}$ and GT$_{+}$ direction respectively. In
Fig.~1 the abscissa shows the energy in the daughter $^{66-71}$Cu
nuclides, while in Fig.~2 the abscissa denotes the energy in the
daughter $^{66-71}$Co nuclides. The charge-changing transition are
shown up to energy of 15 MeV in daughter nuclides. Experimental data
was inserted in the calculation whenever probable. When the computed
excitation energies were in the range of 0.5 MeV of one another then
they were substituted with the experimental levels. The missing
experimental levels were also introduced. The energy levels for
which measured data had no particular parity and/or spin allocation,
and beyond, were not substituted with the theoretical ones. It is to
be noted that in this project forbidden transitions were not
considered. We hope to calculate these forbidden transitions in near
future. It is clear from Figs.~1~and~2 that the charge-changing
transitions in daughter nuclides are well fragmented. It was
observed that the calculated excited states GT strength
distributions were changed from the ground-level strength, which
emphasizes that the Brink-Axel hypothesis is poor estimation to be
used in the computation of astrophysical weak-decay processes. For
nickel isotopes these excited states contribute significantly in the
stellar weak-decay rates during the core contraction and collapse
stages of high mass stars.

The energy dependence of GT strength is not known for several nuclei
of potential significance in presupernova and collapsing stars. The
centroid ($\bar{E}$) of GT strength distribution determine the
effective energy of capturing electron or positron from the ground
level of parent nuclide to excited level of the daughter nuclide.
The weak interaction rates are sensitive exponentially to the
position of GT resonance. Table~2 shows our calculated centroid
($\bar{E}$) and width for the calculated GT distributions along both
$\beta$-decay and EC directions. These computed results are
presented up to cutoff energy (E$_{cut}$) equal to 15 MeV in
daughter nuclides. The values of $\bar{E}$ and widths are displayed
in units of MeV. The pn-QRPA computed $\beta$-decay half-lives
(T$_{1/2}$) were also compared with the experimental half-lives
\cite{aud12}. Table~3 shows a very decent comparison of our
calculated terrestrial T$_{1/2}$ values with the measured data. The
last column in Tab.~3 shows the nuclear deformation ($\beta_{2}$)
values used in this work, obtained by using Eq.~(3). As discussed
earlier these isotopes of nickel are essentially spherical in
nature.

The calculated EC and PC rates for  $^{66-71}$Ni isotopes, as a
function of astrophysical temperature, are displayed in Figs.~3 and
4, respectively. T$_9$ denotes the temperature in unit of 10$^{9}$
K. We present these captures rates at selected densities of 10$^{3}$
g/cm$^{3}$, 10$^{7}$ g/cm$^{3}$ and 10$^{11}$ g/cm$^{3}$,
corresponding to low, medium and high densities, respectively.  The
EC rates are on the rise as the stellar core density and temperature
increase as can be seen from Fig.~3. This is because the increase in
the stellar core density, rise the  Fermi energy of the electrons,
due to which enhancement in EC rates occur. In addition, as the
stellar temperature soars the occupation probability of parent
excited levels increases and contribute effectively to the total
stellar weak-decay rates (Eq.~21). For all these neutron-rich nickel
isotopes, the calculated positron emission rates are negligible and
in the simulation codes these rates might be ignored when matched
with the corresponding EC rates. The effects of PC for the stars
having masses in the range (10 $\leq$ M$_{\odot}$ $\leq$ 40) are
estimated to be smaller and calculated to be bigger for more massive
stars \cite{Heg01}. PC are considered to play a decisive role in the
stellar evolution. These rates decrease the degeneracy pressure of
the electrons available in the astral core. It is noted that the PC
rates are greater than the competing electron emission rates at high
temperatures for all Ni isotopes. The PC rates for the selected Ni
isotopes are shown in Fig.~4. It is observed that the PC rates
increases as the stellar core temperature rises. The PC rates are
similar for the densities in the range (10 -- 10$^{6}$) gcm$^{-3}$.
When the densities of the core increases beyond this range,
decrement in the PC rate starts. In contrast to the EC rates, the PC
rates reduce as the density of the core increases. As temperature
increases or density decreases (at this stage for positrons the
degeneracy parameter become negative), progressively positrons
having very high-energy are generated which results in enhancement
of PC rates. The pn-QRPA computed EC and PC rates on $^{66-71}$Ni
isotopes are presented on selected density-temperature scale in
Tables~4 and 5. We present these capture rates for temperatures in
the range T$_9$ = 1 -- 30 and at selected densities (10$^{2}$,
10$^{5}$, 10$^{8}$ and 10$^{11}$)g/cm$^{3}$. The calculated capture
rates (Eq.~(21)) are displayed in log scale to the base 10. Tables~4
and 5 show that the capture rates increase as stellar temperature
rises. At lower stellar temperature and density the EC rates are
much smaller as compared to PC rates and can safely be neglected.
But in high temperature and low density regions, EC rates compete
rather well with the PC rates. It can also be seen from Tables~4 and
5 that in high stellar densities domain the PC rates are negligible
as compared to EC rates, specially at low T$_9$.

We finally present the comparison of pn-QRPA calculated EC and PC
rates on $^{66-71}$Ni with the earlier calculation of Pruet and
Fuller (hereafter PF) \cite{Pru03}. The strategy adopted by PF for
calculation of lepton capture rates was essentially the same as
adopted by FFN (they also used independent particle model). The weak
rates were split into two portions: a lower portion involving of
discrete transitions amongst individual states whiles the high
portion containing the Fermi and GT resonances. PF  used a
sophisticated treatment of nuclear partition functions in their
calculation.

Figs.~5 and 6 show the comparison of our estimated EC rates with
those computed by PF. In comparing the EC rates, we illustrate three
panels for each isotope of Ni. The upper panel is at stellar density
10 gcm$^{-3}$, the middle  at 10$^{6}$ gcm$^{-3}$ and bottom panel
at 10$^{11}$ gcm$^{-3}$, respectively. These values roughly depict
low, medium and high stellar densities, respectively. Figs.~5 and 6
display the comparison for $^{66,67}$Ni and $^{68-71}$Ni,
respectively. It is observed that at low and medium densities and
high temperatures, our EC rates values are larger than the
corresponding PF calculated rates by up to two orders of magnitude.
In high density and low temperature regions our calculated lepton
capturing rates are in decent comparison with PF, excluding the case
of $^{68}$Ni where PF calculated EC values are an order of magnitude
bigger than the pn-QRPA values. At T$_9$ = 30 (where the probability
of parent excited levels increases considerably) our EC rates values
are greater by an order of magnitude from the corresponding rates of
PF.

The comparison between our calculated PC rates  and those calculated
by PF is presented in Fig.~7. Here we display the ratio of pn-QRPA
computed PC rates to those computed by PF, R$_{PC}$(pn-QRPA/PF). It
is noted that for $^{66-70}$Ni, in low temperature domain, our rates
are in good comparison with PF calculation. At high temperatures our
computed rates are bigger by more than an order of magnitude. For
$^{71}$Ni, even in low temperature regions, our calculated PC rates
values are greater by almost an order of magnitude. The enhancement
of our lepton capture rates have two main reasons. Primarily, a
large model space was used in pn-QRPA calculations (up to seven
major oscillator shells). Further our pn-QRPA model did not employ
Brink-Axel hypothesis and depict a more realistic estimate of weak
rates in stellar matter.  The core-collapse simulators should take
note of our enhanced lepton capture rates on neutron-rich nickel
isotopes.

Since both the PC and electron emission (EE) rates tend to increase
the Y$_{e}$ of stellar matter, one interesting query could be: how
the calculated PC rates compete with the corresponding EE rates for
these  nickel isotopes  in stellar core. Fig.~8 shows the percentage
contribution of PC and EE to total weak rates. The upper two bar
graphs (a) and (b) are shown at low T$_9$ = 2 and high T$_9$ = 30
temperatures, respectively, keeping the stellar density fixed at
$\rho$Y$_{e}$ = 10$^{6}$ g/cm$^{3}$. The bar graphs (c) and (d) are
shown for low stellar density 10 g/cm$^{3}$ and high density
10$^{11}$ g/cm$^{3}$, respectively, keeping  the stellar temperature
constant at T$_9$ = 10. From Fig.~8 it is clear that the
contribution of PC to total rate is negligible at high density and
low temperature. The contribution of PC rate to total weak rate is
large in high temperature regions. At T$_9$=10 and $\rho$Y$_{e}$ =
10 g/cm$^{3}$, both the PC and EE rates compete well with each
other. One may conclude that in high density and low temperature
regions, the contribution of PC values are much smaller as compared
to EE values and may be neglected. However at high temperatures
(late phases of massive stars evolution) the PC rates are the
dominant mode for the stellar weak rates.

\section{Conclusions}
A microscopic and reliable computation of weak-decay processes can
help us  better understand a supernova explosion processes. The
deformed pn-QRPA theory having the good track record of computing
global half-lives values was employed to estimate the lepton
(electron/positron) captures rates on astrophysically significant
neutron-rich isotopes of Ni in stellar environment. The Ikeda sum
rule, which is considered to be model independent, was fulfilled
well in our calculation. The GT strength distributions, centroids
and widths for all nickel isotopes, in both $\beta$-decay and EC
directions, were also computed. The calculated ground as well as
excited states charge-changing transitions were then used for the
computation of total weak-decay rates over wide-ranging stellar
temperatures( T$_9$ = 0.01 -- 30 )and densities (10 -- 10$^{11}$
gcm$^{3}$). We have noted that for these neutron-rich Ni isotopes,
the EC rates are negligible as compared to PC rates in the domains
of low temperature and density values, while in high density regions
(10$^{11}$ g/cm$^{3}$) the PC rates may safely be neglected. In low
to medium stellar density and high temperature regions the EC rates
compete well with the calculated PC rates.

The deformed pn-QRPA calculated capture rates were also compared
with previous calculation of Pruet and Fuller (PF). In high
temperature and low density regions, our EC rates were enhanced by
as much as two orders of magnitude. In low temperature and high
density regions the pn-QRPA calculated rates were in decent matching
with the corresponding PF results (except for the case of
$^{68}$Ni). In high temperature regions our calculated lepton
captures rates  were greater up to an order of magnitude. One of the
main reasons for the big differences in calculated capture rates is
that we computed excited states charge-changing transition in a
microscopic style without assuming the Brink-Axel hypothesis (used
by PF). The deformed pn-QRPA model having a schematic and separable
interaction offered the liberty of having a large model space equal
to 7 $\hbar \omega$ suitable for the treatment of excited levels in
heavier nuclei inside the stellar matter. Another reason for the
differences would be the placement of centroid in the PF and
reported calculations. From astral viewpoint our enhanced lepton
capture rates may have impact on the stellar evolution late stages
and the shock waves energetics. Simulation result shows that the
lepton captures rates have a solid effect on the core collapse path
as well as on the features of the core at bounce. It was also noted
that in high density and low temperature regions, the PC rate
contributions were much smaller as compared to electron emission
(EE) rates. However at elevated temperatures the EE rates
contribution to total rates can safely be neglected.

\section*{Acknowledgements}

J.-U. Nabi would like to acknowledge the support of the Higher
Education Commission Pakistan through the HEC Project No. 20-3099.

\begin{figure}[th]
\centerline{\includegraphics[scale=0.52]{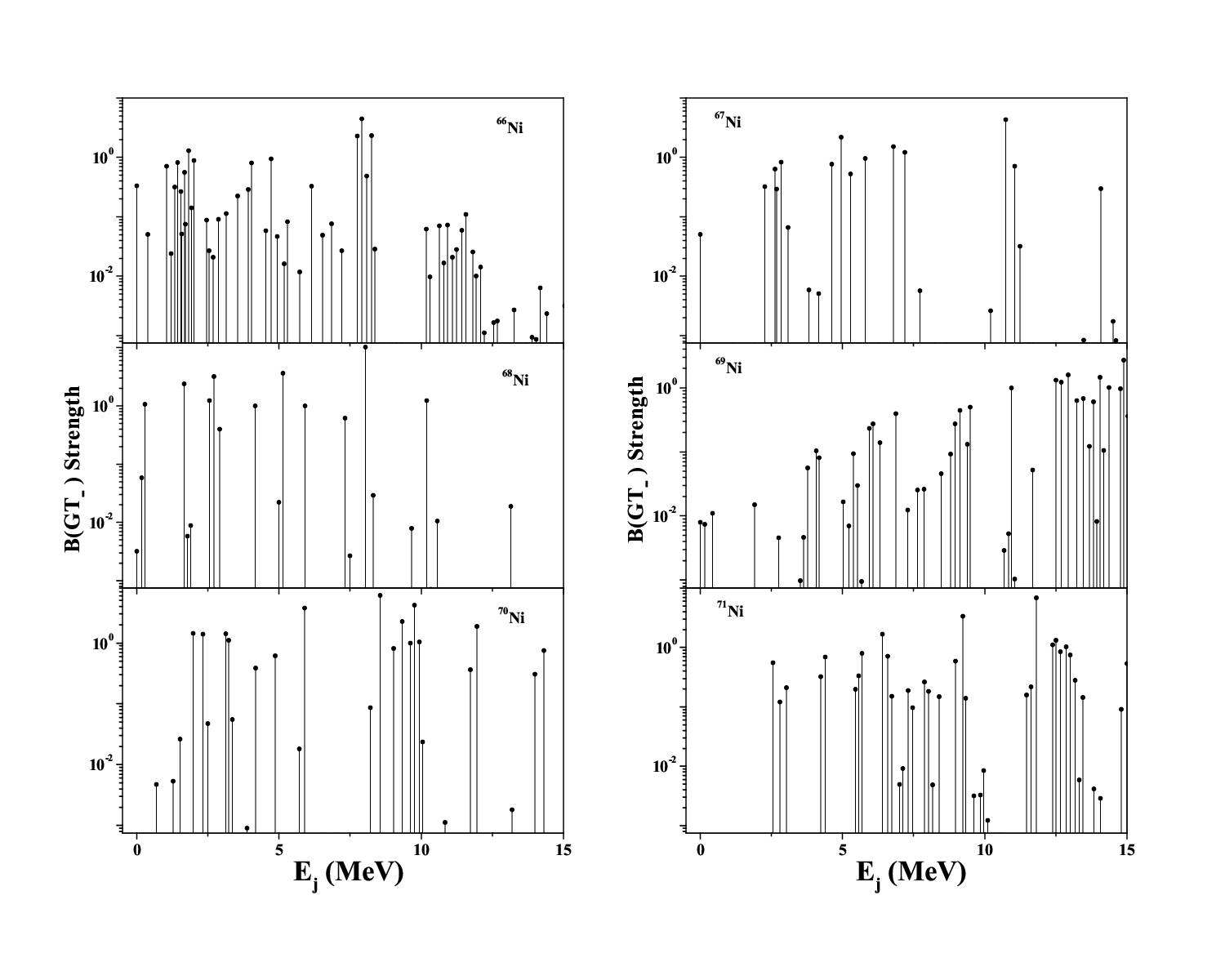}}
\caption{Calculated Gamow-Teller transitions as a function of
daughter excitation energy (E$_j$),  in positron capture direction
using the pn-QRPA model.} \label{fig1}
\end{figure}
\begin{figure}[th]
\centerline{\includegraphics[scale=0.52]{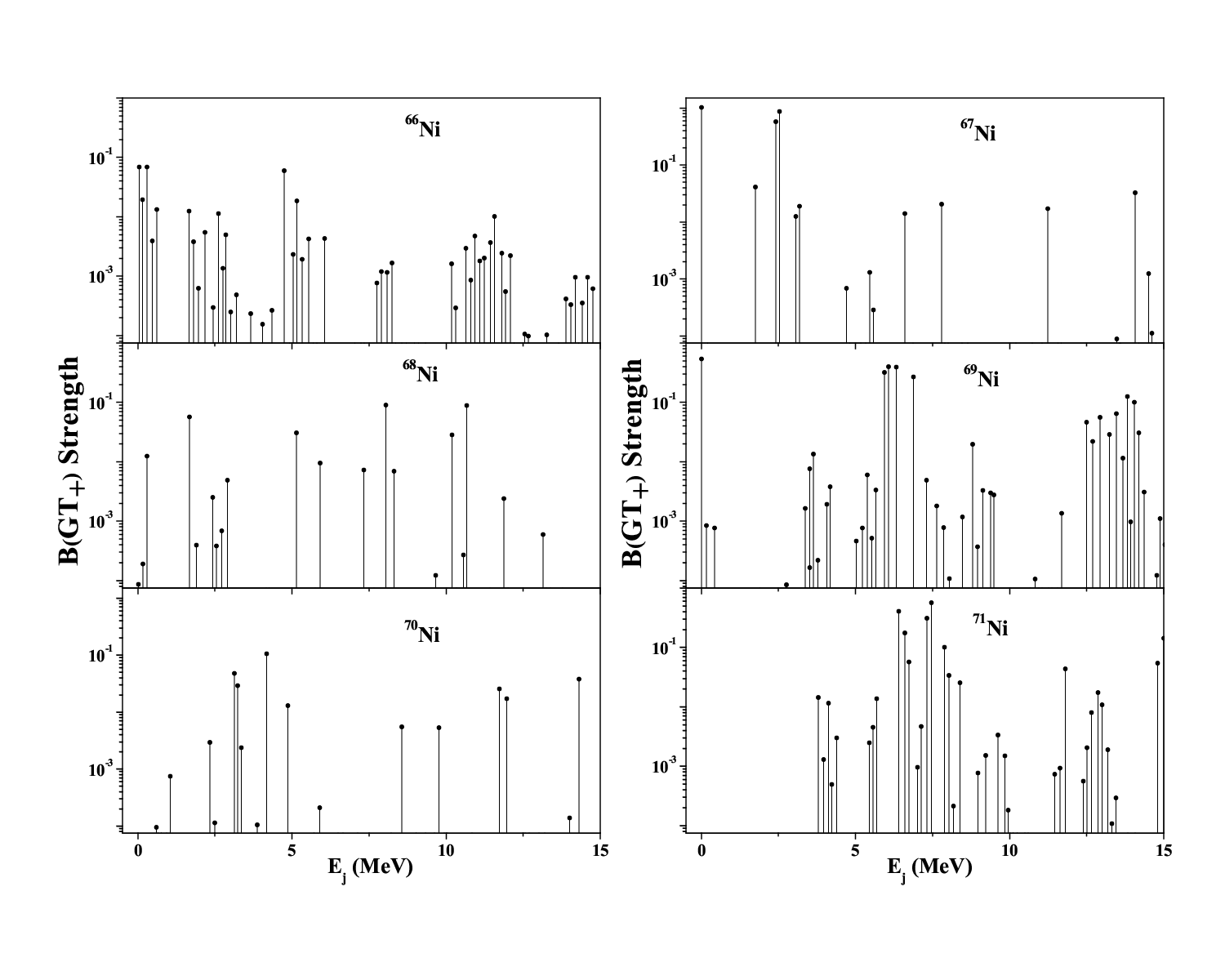}}
\caption{Calculated Gamow-Teller transitions as a function of
daughter excitation energy (E$_j$),  in electron capture direction
using the pn-QRPA model.} \label{fig2}
\end{figure}

\begin{figure}[h]
\centerline{\includegraphics[scale=0.49]{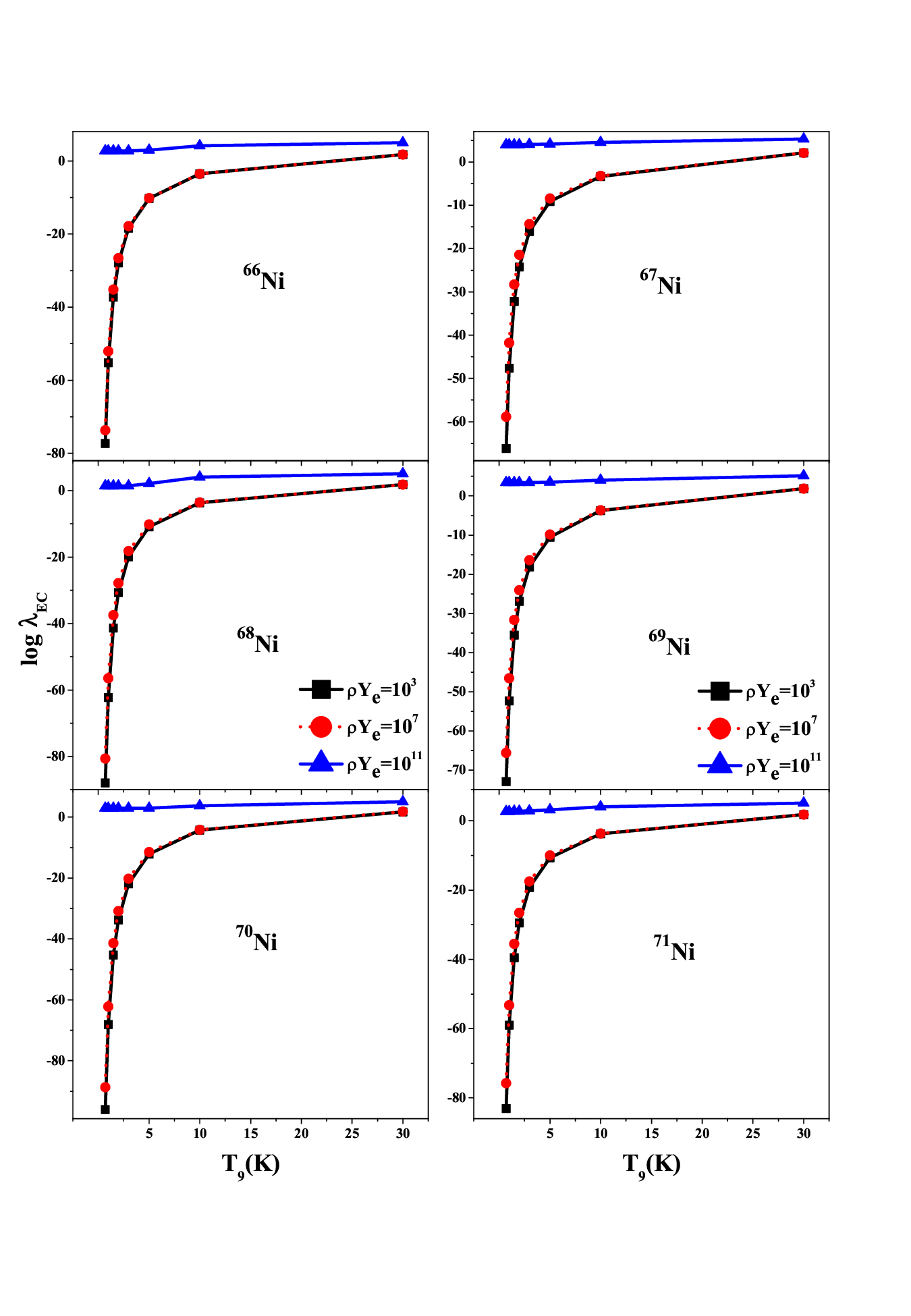}} \caption{The
pn-QRPA calculated EC rates on ($^{66-71}$Ni), as a function of
stellar temperature, for different selected densities. Temperatures
(T$_{9}$) are given in units of 10$^{9}$ K. Stellar densities
($\rho$Y$_{e}$) are given in units of g/cm$^{3}$ and
$\log\lambda_{EC}$ represents the log (to base 10) of EC rates in
units of s$^{-1}$.} \label{fig3}
\end{figure}

\begin{figure}[th]
\centerline{\includegraphics[scale=0.31]{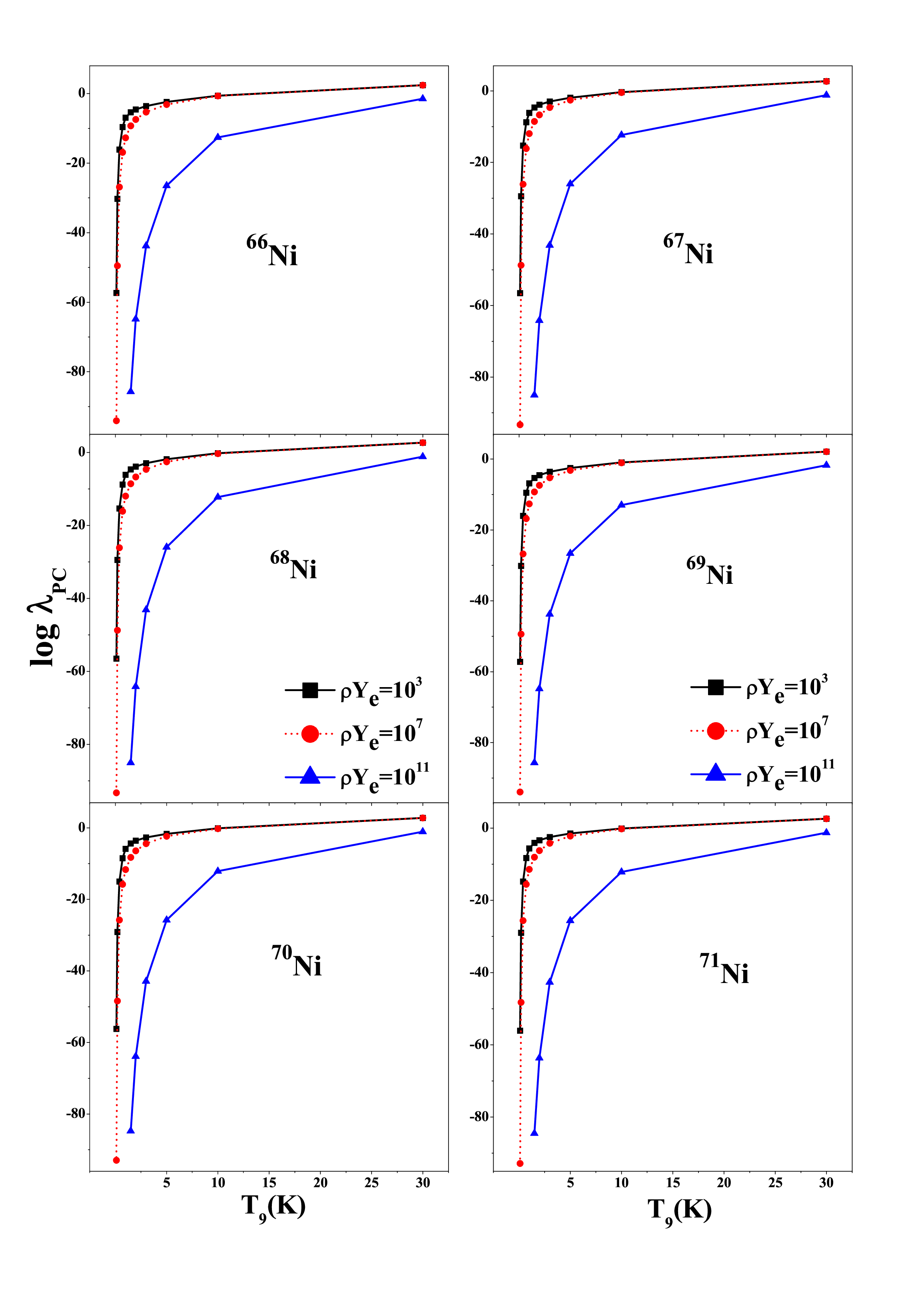}} \caption{The
pn-QRPA calculated PC rates on ($^{66-71}$Ni), as a function of
stellar temperature, for different selected densities. Temperature
(T$_{9}$) is given in units of 10$^{9}$ K. Stellar density
($\rho$Y$_{e}$) is given in units of g/cm$^{3}$ and
$\log\lambda_{PC}$ represents the log (to base 10) of PC rates in
units of s$^{-1}$.}\label{fig4}
\end{figure}

\begin{figure}[th]
\centerline{\includegraphics[scale=0.31]{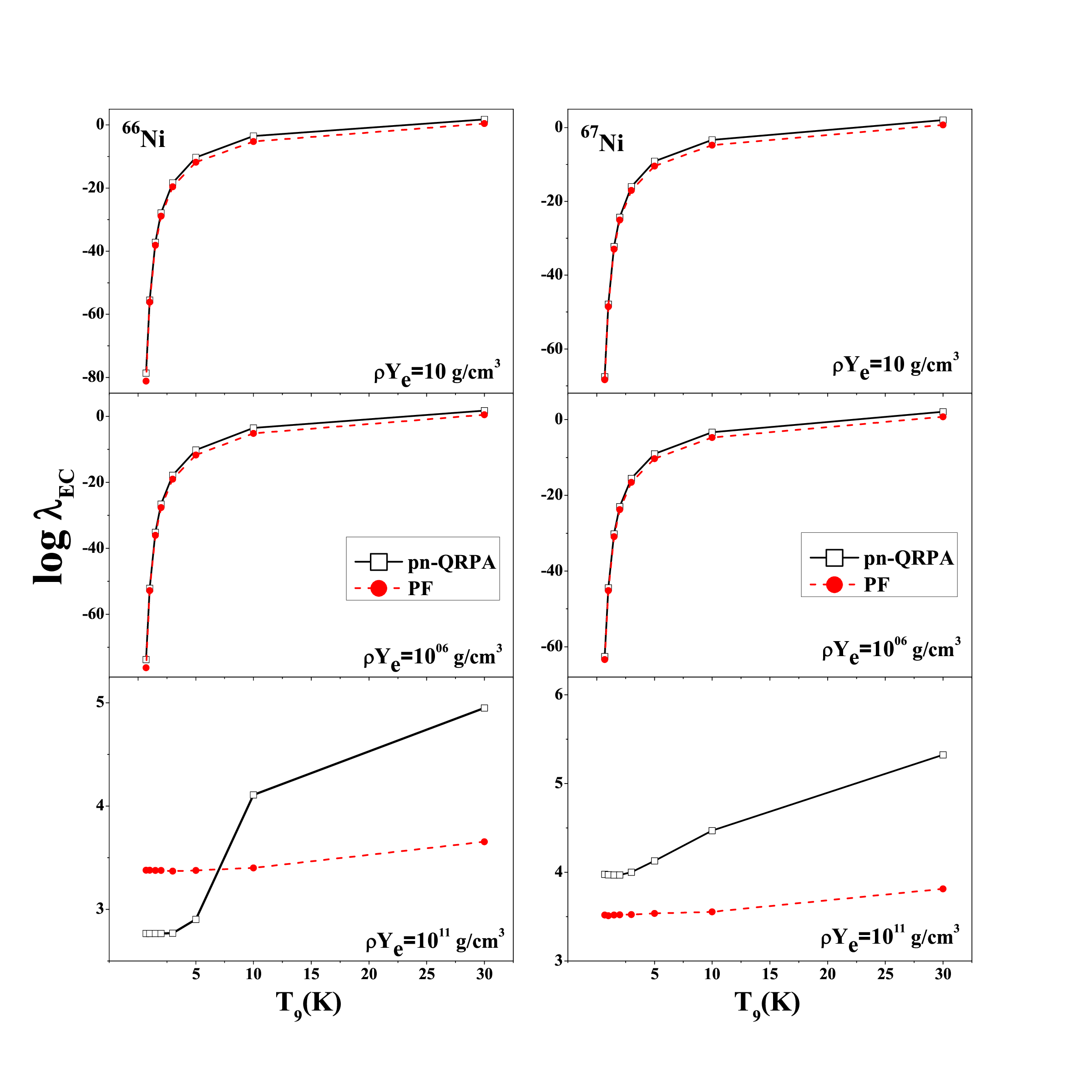}}
\caption{Comparison of pn-QRPA calculated EC rates  (this work) with
those calculated by PF \cite{Pru03}. Temperature (T$_{9}$) is given
in units of 10$^{9}$ K. Stellar density ($\rho$Y$_{e}$) is given in
units of g/cm$^{3}$ and $\log \lambda_{EC}$ represents the log (to
base 10) of EC rates in units of s$^{-1}$.}\label{fig5}
\end{figure}

\begin{figure}[h]
\centerline{\includegraphics[scale=0.31]{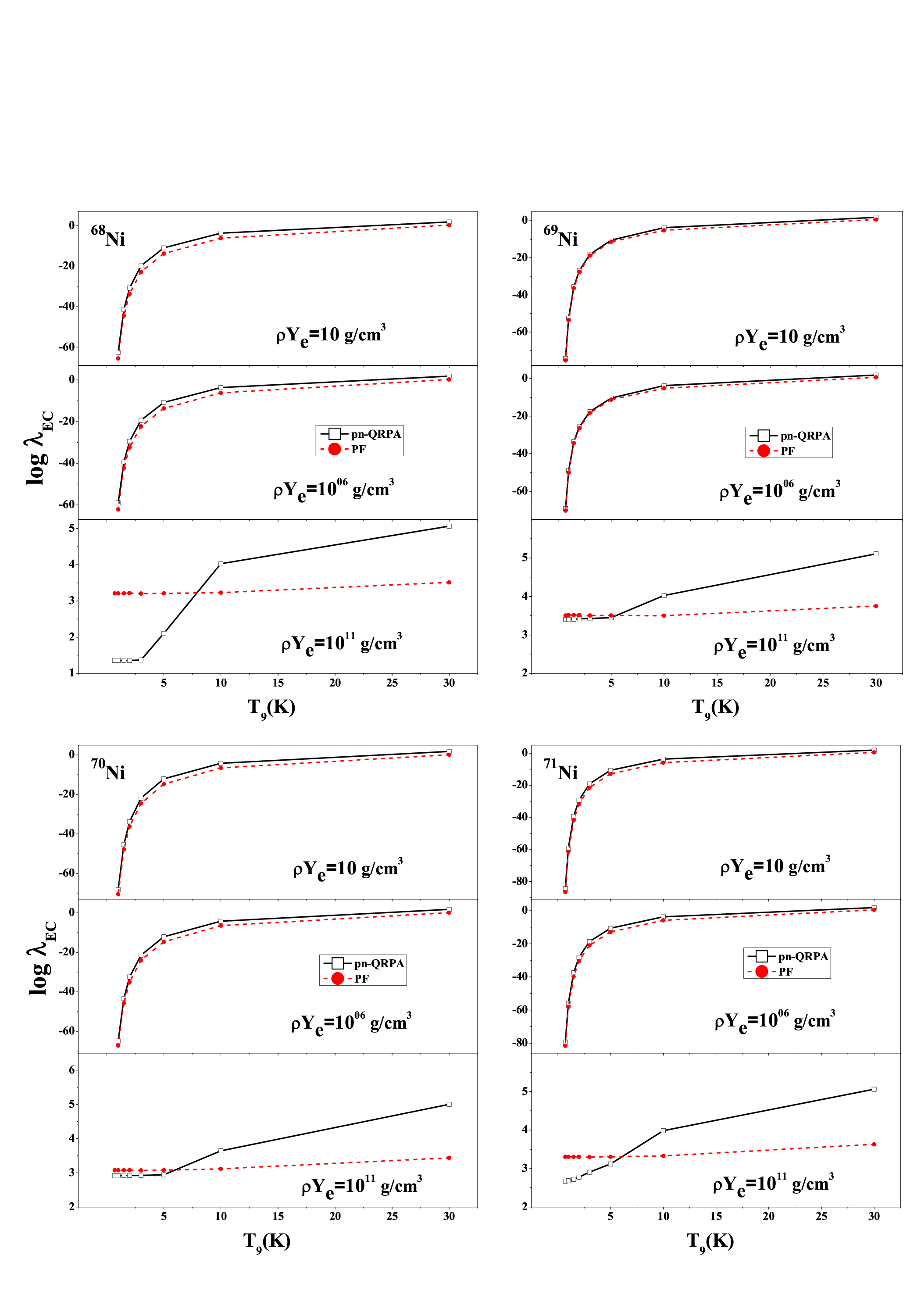}}
\caption{Same as
Fig. 5, but for $^{68-71}$Ni.}\label{fig6}
\end{figure}

\begin{figure}[th]
\centerline{\includegraphics[scale=0.31]{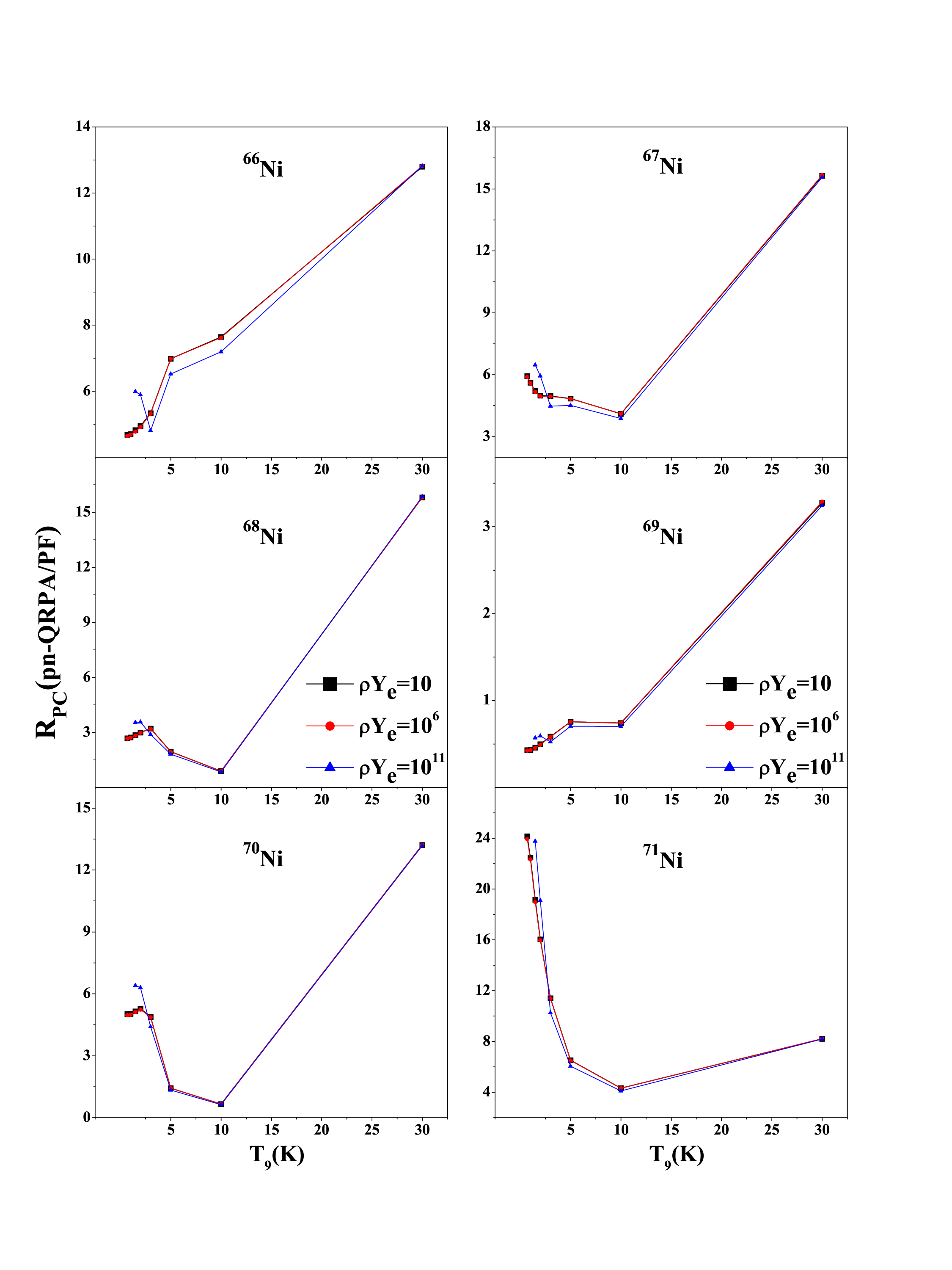}} \caption{Ratio
of reported PC rates  to those calculated by PF \cite{Pru03} as a
function of stellar temperature, for different selected densities.
Temperature (T$_{9}$) is given in units of 10$^{9}$ K and stellar
density ($\rho$Y$_{e}$) is given in units of
g/cm$^{3}$.}\label{fig7}
\end{figure}

\begin{figure}[th]
\centerline{\includegraphics[scale=0.3]{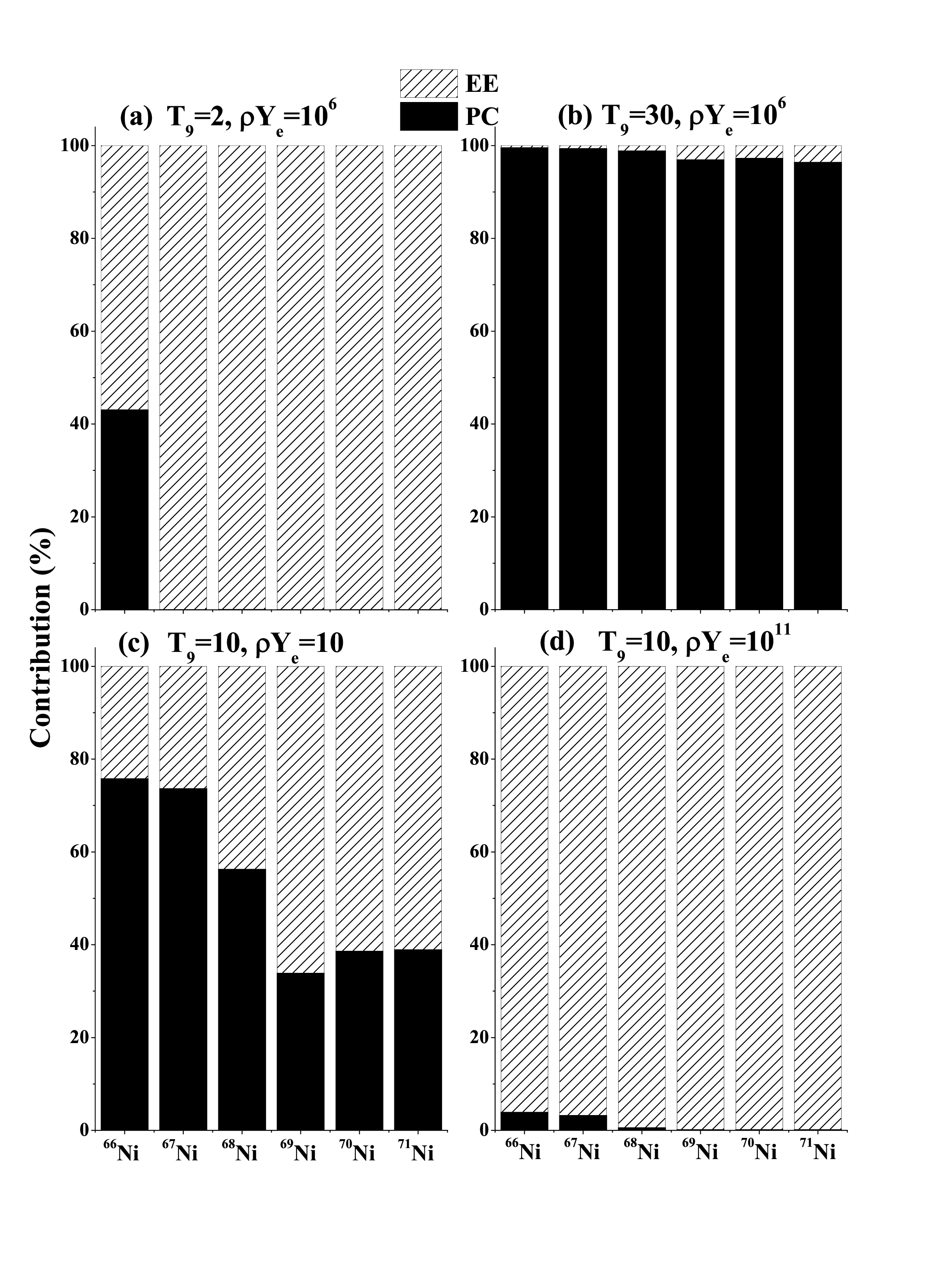}}
\caption{Percentage contribution of PC and electron emission (EE)
rates to total weak rate. Temperature (T$_{9}$) is given in units of
10$^{9}$ K. Stellar density ($\rho$Y$_{e}$) is given in units of
g/cm$^{3}$.} \label{fig8}
\end{figure}

\clearpage
\begin{table}[pt]
  \centering
\scriptsize  \caption{Comparison of (re-normalized) calculated and
theoretical Ikeda sum rule for $^{66-71}$Ni.}\label{Table 1}
    \begin{tabular}{ccccccc}
\hline
A & N & $\sum$B(GT$_{-}$)& $\sum$B(GT$_{+}$)& Re-ISR$_{cal}$& Re-ISR$_{th}$ & Cutoff energy in daughter (MeV)\\
\hline
66    & 38    & 11.11 & 0.31  & 10.80 & 10.80 & 7.76 \\
67    & 39    & 14.45 & 2.59  & 11.87 & 11.88 & 13.59 \\
68    & 40    & 13.09 & 0.13  & 12.96 & 12.96 & 7.32 \\
69    & 41    & 16.52 & 2.49  & 14.02 & 14.04 & 14.88 \\
70    & 42    & 15.32 & 0.21  & 15.12 & 15.12 & 8.56 \\
71    & 43    & 17.98 & 1.78  & 16.19 & 16.20 & 11.81 \\
\end{tabular}
\end{table}

\begin{table}[pt]
  \centering
\scriptsize  \caption{Calculated centroid  and width of GT
distributions of Ni isotopes in $\beta$-decay and EC directions. The
cutoff energy in daughter nuclei is 15 MeV.}\label{Table 2}
    \begin{tabular}{ccccc}
\hline
    Nuclei & $\bar{E}_{-}$ (MeV)& $\bar{E}_{+}$ (MeV)&Width$_{-}$ (MeV)&Width$_{+}$ (MeV) \\
    \hline
$^{66}$Ni & 5.56  & 2.90  & 3.01  & 3.48 \\
$^{67}$Ni & 7.21  & 1.77  & 3.15  & 2.09 \\
$^{68}$Ni & 5.57  & 7.13  & 2.91  & 3.47 \\
$^{69}$Ni & 12.50 & 6.35  & 2.70  & 4.39 \\
$^{70}$Ni & 7.80  & 6.53  & 3.18  & 4.20 \\
$^{71}$Ni & 9.95  & 8.03  & 3.04  & 2.62 \\

\end{tabular}
\end{table}

\begin{table}[pt]
  \centering
\scriptsize \caption{The pn-QRPA calculated $\beta$-decay half-lives
compared with the measured ones \cite{aud12}. The last column shows
the nuclear deformation ($\beta_{2}$) values used in our
calculation.}\label{Table 3}
    \begin{tabular}{cccc}
\hline
Nuclei & T$_{1/2}$(cal) (s)& T$_{1/2}$(exp) (s)& $\beta_{2}$\\
\hline
$^{66}$Ni & 210778.1 & 196560.0  &  -0.034 \\
$^{67}$Ni & 21.9     & 21.0  &  0.004 \\
$^{68}$Ni & 30.6     & 29.0  &  -0.004 \\
$^{69}$Ni & 11.5     & 11.5  &  -0.028 \\
$^{70}$Ni & 6.1      & 6.0  & -0.004 \\
$^{71}$Ni & 2.7      & 2.6  &  0.016 \\

\end{tabular}
\end{table}

\begin{table}[pt]
\centering \scriptsize \caption{Calculated electron capture (EC) and
positron capture (PC) rates on $^{66,67,68}$Ni isotopes in stellar
matter. The first column shows the stellar density ($\rho$Y$_{e}$)
(in units of gcm$^{-3}$). Temperature (T$_{9}$) is given in units of
10$^{9}$ K. The calculated EC and PC rates are tabulated in
logarithmic (to base 10) scale in units of s$^{-1}$. In the table,
-100.00 means that the rate is smaller than
10$^{-100}$.}\label{Table 4}
    \begin{tabular}{|c|c|cc|cc|cc|}

$\rho$$\it Y_{e}$ & T$_{9}$  & \multicolumn{2}{c|}{$^{66}$Ni}& \multicolumn{2}{c|}{$^{67}$Ni}& \multicolumn{2}{c|}{$^{68}$Ni} \\
\cline{3-8} & &EC & PC& EC& PC& EC& PC\\
\hline
10$^{2}$     & 1     & -55.47 & -6.69 & -47.82 & -5.93 & -62.51 & -5.92 \\
10$^{2}$     & 1.5   & -37.21 & -5.34 & -32.25 & -4.62 & -41.43 & -4.59 \\
10$^{2}$     & 2     & -27.91 & -4.56 & -24.28 & -3.88 & -30.75 & -3.83 \\
10$^{2}$     & 3     & -18.37 & -3.58 & -16.06 & -2.96 & -19.90 & -2.90 \\
10$^{2}$     & 5     & -10.30 & -2.41 & -9.15 & -1.90 & -10.92 & -1.84 \\
10$^{2}$     & 10    & -3.53 & -0.59 & -3.34 & -0.35 & -3.69 & -0.21 \\
10$^{2}$     & 30    & 1.72  & 2.42  & 2.07  & 2.68  & 1.82  & 2.71 \\
\hline
10$^{5}$     & 1     & -53.41 & -8.75 & -45.76 & -7.99 & -60.43 & -7.98 \\
10$^{5}$     & 1.5   & -36.32 & -6.23 & -31.36 & -5.51 & -40.54 & -5.48 \\
10$^{5}$     & 2     & -27.59 & -4.88 & -23.96 & -4.20 & -30.43 & -4.15 \\
10$^{5}$     & 3     & -18.31 & -3.64 & -16.00 & -3.02 & -19.84 & -2.96 \\
10$^{5}$     & 5     & -10.29 & -2.42 & -9.14 & -1.90 & -10.91 & -1.84 \\
10$^{5}$     & 10    & -3.53 & -0.59 & -3.34 & -0.35 & -3.69 & -0.21 \\
10$^{5}$     & 30    & 1.72  & 2.42  & 2.07  & 2.68  & 1.82  & 2.71 \\
\hline
10$^{8}$     & 1     & -43.25 & -18.95 & -35.57 & -18.19 & -50.23 & -18.17 \\
10$^{8}$     & 1.5   & -29.09 & -13.48 & -24.11 & -12.76 & -33.28 & -12.73 \\
10$^{8}$     & 2     & -21.86 & -10.62 & -18.22 & -9.93 & -24.69 & -9.89 \\
10$^{8}$     & 3     & -14.42 & -7.53 & -12.10 & -6.91 & -15.94 & -6.85 \\
10$^{8}$     & 5     & -8.09 & -4.61 & -6.94 & -4.09 & -8.71 & -4.03 \\
10$^{8}$     & 10    & -2.78 & -1.33 & -2.59 & -1.09 & -2.94 & -0.95 \\
10$^{8}$     & 30    & 1.76  & 2.39  & 2.10  & 2.65  & 1.86  & 2.68 \\
\hline
10$^{11}$    & 1     & 2.77  & -100.00 & 3.97  & -100.00 & 1.36  & -100.00 \\
10$^{11}$    & 1.5   & 2.77  & -85.75 & 3.97  & -85.03 & 1.36  & -85.00 \\
10$^{11}$    & 2     & 2.77  & -64.86 & 3.97  & -64.18 & 1.36  & -64.13 \\
10$^{11}$    & 3     & 2.77  & -43.77 & 4.00  & -43.15 & 1.37  & -43.08 \\
10$^{11}$    & 5     & 2.90  & -26.50 & 4.13  & -25.98 & 2.10  & -25.92 \\
10$^{11}$    & 10    & 4.11  & -12.59 & 4.47  & -12.35 & 4.03  & -12.21 \\
10$^{11}$    & 30    & 4.95  & -1.44 & 5.32  & -1.18 & 5.06  & -1.15 \\

\end{tabular}
\end{table}

\begin{table}[pt]
\centering \scriptsize \caption{Same as Table~3 but for
$^{69,70,71}$Ni}\label{Table 5}
    \begin{tabular}{|c|c|cc|cc|cc|}

$\rho$$\it Y_{e}$ & T$_{9}$ & \multicolumn{2}{c|}{$^{69}$Ni}& \multicolumn{2}{c|}{$^{70}$Ni}& \multicolumn{2}{c|}{$^{71}$Ni} \\
\cline{3-8} & &EC & PC& EC& PC& EC& PC\\
\hline
10$^{2}$     & 1     & -52.59 & -6.63 & -68.25 & -5.64 & -59.25 & -5.41 \\
10$^{2}$     & 1.5   & -35.59 & -5.30 & -45.35 & -4.33 & -39.49 & -4.11 \\
10$^{2}$     & 2     & -26.96 & -4.53 & -33.74 & -3.59 & -29.47 & -3.37 \\
10$^{2}$     & 3     & -18.14 & -3.59 & -21.93 & -2.69 & -19.23 & -2.49 \\
10$^{2}$     & 5     & -10.55 & -2.51 & -12.16 & -1.69 & -10.70 & -1.51 \\
10$^{2}$     & 10    & -3.76 & -0.96 & -4.25 & -0.14 & -3.74 & -0.14 \\
10$^{2}$     & 30    & 1.86  & 2.13  & 1.74  & 2.75  & 1.82  & 2.59 \\
\hline
10$^{5}$     & 1     & -50.53 & -7.66 & -66.19 & -7.70 & -57.19 & -7.47 \\
10$^{5}$     & 1.5   & -34.70 & -5.46 & -44.46 & -5.22 & -38.60 & -5.00 \\
10$^{5}$     & 2     & -26.64 & -4.57 & -33.41 & -3.91 & -29.14 & -3.69 \\
10$^{5}$     & 3     & -18.08 & -3.59 & -21.87 & -2.75 & -19.17 & -2.55 \\
10$^{5}$     & 5     & -10.54 & -2.51 & -12.15 & -1.70 & -10.69 & -1.52 \\
10$^{5}$     & 10    & -3.76 & -0.96 & -4.25 & -0.15 & -3.74 & -0.14 \\
10$^{5}$     & 30    & 1.86  & 2.13  & 1.74  & 2.75  & 1.82  & 2.59 \\
\hline
10$^{8}$     & 1     & -40.33 & -18.89 & -55.99 & -17.90 & -46.99 & -17.67 \\
10$^{8}$     & 1.5   & -27.45 & -13.45 & -37.20 & -12.47 & -31.35 & -12.25 \\
10$^{8}$     & 2     & -20.90 & -10.59 & -27.68 & -9.65 & -23.40 & -9.43 \\
10$^{8}$     & 3     & -14.18 & -7.54 & -17.97 & -6.64 & -15.28 & -6.44 \\
10$^{8}$     & 5     & -8.34 & -4.70 & -9.95 & -3.88 & -8.49 & -3.71 \\
10$^{8}$     & 10    & -3.01 & -1.70 & -3.50 & -0.88 & -2.99 & -0.88 \\
10$^{8}$     & 30    & 1.90  & 2.09  & 1.78  & 2.72  & 1.86  & 2.56 \\
\hline
10$^{11}$    & 1     & 3.40  & -100.00 & 2.92  & -100.00 & 2.68  & -100.00 \\
10$^{11}$    & 1.5   & 3.41  & -85.71 & 2.92  & -84.74 & 2.72  & -84.51 \\
10$^{11}$    & 2     & 3.42  & -64.83 & 2.92  & -63.89 & 2.78  & -63.67 \\
10$^{11}$    & 3     & 3.42  & -43.77 & 2.92  & -42.88 & 2.91  & -42.67 \\
10$^{11}$    & 5     & 3.45  & -26.59 & 2.95  & -25.77 & 3.12  & -25.60 \\
10$^{11}$    & 10    & 4.02  & -12.95 & 3.64  & -12.14 & 3.98  & -12.13 \\
10$^{11}$    & 30    & 5.11  & -1.73 & 5.00  & -1.10 & 5.06  & -1.26 \\

\end{tabular}
\end{table}

\end{document}